\documentclass[12pt]{article}

\usepackage{jheppub}
\usepackage{mathrsfs}
\usepackage{bbm}
\usepackage{bbold}
\usepackage{slashed}

\newcommand{\be}{\begin{equation}}
\newcommand{\ee}{\end{equation}}
\newcommand{\bi}{\begin{itemize}}
\newcommand{\ei}{\end{itemize}}
\newcommand{\bea}{\begin{eqnarray}}
\newcommand{\eea}{\end{eqnarray}}

\newcommand{\Nabla}{\boldsymbol{\nabla}}

\def\lambdabar{\protect\@lambdabar}
\def\@lambdabar{%
\relax \bgroup
\def\@tempa{\hbox{\raise.73\ht0
\hbox to0pt{\kern.2\wd0\vrule width.7\wd0
height.1pt depth.1pt\hss}\box0}}%
\mathchoice{\setbox0\hbox{$\displaystyle\lambda$}\@tempa}%
{\setbox0\hbox{$\textstyle\lambda$}\@tempa}%
{\setbox0\hbox{$\scriptstyle\lambda$}\@tempa}%
{\setbox0\hbox{$\scriptscriptstyle\lambda$}\@tempa}%
\egroup }

\newcommand{\vc}[1]{\mbox{\textbf{#1}}}

\newcommand{\pad}[2]{\frac{\partial #1}{\partial #2}}
\newcommand{\sfrac}[2]{{\textstyle \frac{#1}{#2}}}

\begin{document}

\title{QED and Lasers: A Tutorial}

\author[1]{T.~Heinzl,}
\affiliation[1]{Centre for Mathematical Sciences, School of Engineering, Computing and Mathematics, University of Plymouth, PL4 8AA, UK}
\emailAdd{theinzl@plymouth.ac.uk}



\abstract{This is a write-up of a short tutorial talk on high-intensity QED, video-presented at the 2021 annual Christmas meeting of the Central Laser Facility at Rutherford-Appleton Lab, UK. The first half consists of a largely historical introduction to (quantum) electrodynamics focussing on a few key concepts. This well-established theory is then compared to its strong-field generalisation when a high-intensity laser is present. Some supplementary material and a fair amount of references have been added.

\vspace{1cm}

\begin{center}
\textit{In memoriam: Ernst Werner (1930-2021), `Doktorvater' and mentor.}
\end{center}
}

\maketitle

\section{Introduction: A Story of Two Acronyms}

The title of this tutorial admittedly sounds a bit mundane -- but then it is a fairly precise description of what we will do: we will rather literally add a very strong laser (field) to quantum electrodynamics (QED), the microscopic theory of light-matter interactions. With the preceding sentence we have defined the first acronym, QED. Its invention has its own complicated history and arguably goes back all the way to Planck's analysis of black-body radiation and his introduction of energy quanta \cite{Planck:1900a}, later to be identified with photons, the particles of light. The quantum field theoretical aspects (`second quantisation') were introduced in the \emph{Dreim\"annerarbeit} by Born, Heisenberg and Jordan \cite{Born:1926}, the latter being responsible for the section named `Coupled harmonic oscillators. Statistics of Wave fields.' In this section, Jordan quantised the one-dimensional string which corresponds to a  scalar, hence spin zero, field.  Quantisation of the \emph{electromagnetic} field, a spin-one vector field, was first achieved by Dirac a year later in 1927 \cite{Dirac:1927a}, which thus may be viewed as the proper year of birth for QED. Useful collections of early papers on QED may be found in the source texts \cite{Schwinger:1958,vanderWaerden:1968,Miller:1994}. The history of QED has been described in the magisterial tome by Schweber \cite{Schweber:1994qa}.

The laser has become a household name, so one tends to forget that it is another acronym standing for `light amplification by the stimulated emission of radiation'. Stimulated emission is a quantum process that was first explained by Einstein in terms of his eponymous coefficients \cite{Einstein:1916}. Enhancing this effect employing population inversion and a gain medium was achieved by 1960 when the first laser was built  by Maiman \cite{Maiman:1960}. A particularly important breakthrough in our context was achieved in 1985 with the invention of chirped pulse amplification (CPA) by Strickland and Mourou \cite{Strickland:1985gxr}, which earned them the Nobel prize in 2018. For the purposes of this tutorial, we will take such a CPA based high-intensity laser for granted and discuss how its presence affects the fundamental processes of QED. More details on the fundamentals and applications of lasers may be found in the comprehensive texts \cite{Siegman:1986,Svelto:2010,Meschede:2004}.  

Incidentally, the births of QED and the laser are related theoretically: Dirac's inaugural 1927 paper \cite{Dirac:1927a} calculates the Einstein coefficients using QED. A modern account  of these calculations can be found at the very beginning of Schwartz's recent text on quantum field theory~\cite{Schwartz:2014sze}.  

\section{Electrodynamics}

The modern theory of electrodynamics came into being with Maxwell's equations as summarised in Maxwell's famous (but now little read) treatise of 1873 \cite{Maxwell:1873}. One of his big achievements was the (theoretical) identification of light as electromagnetic radiation. This finally clarified the earlier observation by Weber \cite{Weber:1846} that electromagnetism seemed to naturally involve a constant of nature with units of speed\footnote{In Weber's (wrong) electrodynamical equation, based on action-at-a-distance rather than the field concept, this speed corresponds to $\sqrt{2}$ times the speed of light, $c$. In an experiment with Kohlrausch \cite{Weber:1855}, they determined the speed implying a value of $c=3.1 \times 10^8$ m/s (if one anachronistically employs modern terminology).}. Hertz's discovery of electromagnetic radiation \cite{Hertz:1892} was the experimental confirmation that finally confirmed Maxwell's prediction. 

Maxwell's treatise made for rather difficult reading as he did not have 3-vector notation at hand. This was only introduced a decade later by Heaviside \cite{Heaviside:1893} who (together with Hertz \cite{Hertz:1890a}) gave Maxwell's equations their modern form\footnote{Nahin in his Heaviside biography \cite{Nahin:2002} states that for a short while the reformulated Maxwell equations were called the `Hertz-Heaviside equations'.}.   Thus, in 1894 Hertz could confidently state that \emph{``Maxwell's theory is the system of  Maxwell's equations''}. Interestingly, Hertz, who died in 1894 aged 36, was not able to solve the problem of the electrodynamics of moving bodies as he based his discussion in \cite{Hertz:1890b} on Galilei symmetry. However, the symmetry transformations of Maxwell's equations are the Lorentz transformations discovered by Lorentz in 1895 \cite{Lorentz:1895}. The situation may be analysed as follows: If Maxwell's equations were Galilei covariant and thus had the same form in all inertial frames related through Galilei transformations, it should be possible to eliminate the speed of light, $c$, from the equations. Choosing Heaviside-Lorentz units and rescaling the magnetic field, $\vc{B} \to c \vc{B}$, while leaving the electric field, $\vc{E}$, untouched, $c$ can indeed be eliminated from all equations except for Faraday's induction law which becomes
\be
  \Nabla \times \vc{E} = -\frac{1}{c^2} \pad{\vc{B}}{t} \; . \label{FARADAY}
\ee
In other words, sending $c$ to infinity, we would end up with a Galilei covariant version of electrodynamics losing the induction law (\ref{FARADAY}) on the way \cite{LeBellac:1973,Jammer:1980}. 

It took an Einstein to finally solve the problem of reconciling motion and electrodynamics in his aptly entitled paper \emph{On the Electrodynamics of Moving Bodies} \cite{Einstein:1905ve}. There, he formulated his two postulates of special relativity, namely that \emph{all} laws of physics take on the same form in inertial frames related by Lorentz transformations (Lorentz covariance) and that the speed of light is universal, i.e.\ it has the same value in all inertial frames (Lorentz invariance). 

The final touch was provided by Minkowski \cite{Minkowski:1908a} who introduced 4-vector notation thus making Lorentz covariance manifest. All one has to do is to find an appropriate `zero component' and `add' it to Heaviside's 3-vectors.  4-positions are \emph{space-time} vectors $x = (ct, \vc{x})$ implying a 4-gradient $\partial = (\partial_{ct}, \Nabla)$ and so on. In doing so, one introduces what is now called `Minkowski space', which Minkowski described as follows: \textsl{The consequences will be radical. Henceforth, `space’ and `time’, viewed as separate entities, will turn into mere shadows entirely, and only a union of the two will retain an independent meaning} \cite{Minkowski:1908b}.

In the context of electrodynamics, one augments the electromagnetic current, $\vc{j}$,  by $j^0 = c\rho$, where $\rho$ is the charge density, implying a 4-current $j = (c\rho, \vc{j})$. Electric and magnetic fields, $\vc{E}$ and $\vc{B}$, do not transform as 3-vectors, but rather as the components of an electromagnetic tensor, $F = F(\vc{E}, \vc{B})$ and its dual, $\tilde{F} = F(\vc{B}, -\vc{E})$. The numbers of degrees of freedom (six) match if $F$ is anti-symmetric, $F^T = -F$. With these ingredients, Maxwell's equations can finally be written in a form that remains unchanged upon changing frames. Thus, all inertial observers will agree on the following equations:
\begin{eqnarray}
  \partial_\mu F^{\mu\nu} &=& j^\nu \; , \\ \label{MAXWELL.INH}
  \partial_\mu \tilde{F}^{\mu\nu} &=& 0 \; . \label{MAXWELL.HOM}
\end{eqnarray}
As important as covariant quantities (4-vectors, tensors, ...) are Lorentz invariant ones, often referred to as scalars. In practical terms, these have fully contracted Lorentz indices, so all of these appear twice and are summed over. Regarding the electromagnetic fields, the most important scalars are
\begin{eqnarray}
  \mathcal{S} &=& - \frac{1}{4} F_{\mu\nu} F^{\mu\nu} = \frac{1}{2} (E^2 - B^2) \; , \label{S} \\
  \mathcal{P} &=& - \frac{1}{4} F_{\mu\nu} \tilde{F}^{\mu\nu} = \vc{E} \cdot \vc{B} \; . \label{P}
\end{eqnarray}
These may in turn be used to define Lorentz invariant field magnitudes, which are basically given in terms of the real eigenvalues of $F$,
\begin{eqnarray}
  \mathcal{E} &=& (\sqrt{\mathcal{S}^2 + \mathcal{P}^2} + \mathcal{S})^{1/2} \; , \label{E.INV} \\
  \mathcal{B} &=& (\sqrt{\mathcal{S}^2 + \mathcal{P}^2} - \mathcal{S})^{1/2} \; . \label{B.INV}
\end{eqnarray}
The scalar quantity $\mathcal{S}$ serves as the  Maxwell Lagrangian in vacuum, i.e.\ in the absence of charges. To couple this to charged matter one needs another 4-vector, the electromagnetic (or gauge) potential which combines the (rotational) scalar and 3-vector potentials according to $A = (\phi, \vc{A})$. The field strength is then its covariant curl, $F_{\mu\nu} = \partial_\mu A_\nu - \partial_\nu A_\mu$. Note that this solves the homogeneous Maxwell equations by construction as 
\be 
  \partial_\mu \tilde{F}^{\mu\nu} = \epsilon^{\mu\nu\rho\sigma} \partial_\mu \partial_\rho A_\sigma = 0 
\ee
by the anti-symmetry of the Levi-Civita tensor $\epsilon$ (or, if you like, the Minkowski version of div curl = 0). The coupling to matter is now given by the Lorentz invariant scalar product of $A$ and the electromagnetic current $j$ which, for a relativistic point particle of mass $m$ and charge $e$ on a space-time curve (world-line) $x = x(\tau)$, takes on the form
\be 
   j^\mu(x) = ec \int d\tau \,  \dot{x}^\mu \, \delta^4 (x - x(\tau)) \; .
\ee
The coupling term in the action  is thus
\be 
  S_\mathrm{int} = \frac{1}{c} \int d^4 x \, j \cdot A = e \int d\tau\,  \dot{x} \cdot A \; .
\ee
The complete Lorentz invariant action for electrodynamics coupled to matter was first written down by Schwarzschild in 1903 \cite{Schwarzschild:1903a} and reads
\be 
  S = \int d\tau \,  (-mc^2 + e \dot{x} \cdot A) + \int d^4 x \, \mathcal{S} \; . \label{S.ED}
\ee
In the above, $cd\tau = ds= (dx \cdot dx)^{1/2}$ denotes the invariant distance element with proper time $\tau$. We mention in passing that the action (\ref{S.ED}) is \emph{gauge invariant}, that is invariant under local transformations $A_\mu \to A_\mu + \partial_\mu \Lambda$, which only add a total time derivative to the Lagrangian, $(\dot{x} \cdot \partial) \Lambda = d\Lambda/d\tau$, and hence leave the equations of motion unchanged\footnote{The notion of gauge invariance goes back to Weyl \cite{Weyl:1918ib,Weyl:1929fm} as nicely reviewed in \cite{ORaifeartaigh:1997dvq,Straumann:2005hj}}. The latter are obtained via Hamilton's principle of least action and read
\begin{eqnarray}
  \Box A_\mu &=& j_\mu \; , \\[5pt] \label{INH.WAVE.EQ}
  m \ddot{x}^\mu &=& \frac{e}{c} F^{\mu\nu} \dot{x}_\nu \; . \label{LORENTZ}
\end{eqnarray}
Equation (\ref{INH.WAVE.EQ}) is the inhomogeneous wave equation which is solved by the Lienard-Wiechert potential, formally $A^\mu = A^\mu_0 + \Box^{-1} j^\mu$, the inverse wave operator representing the retarded Green function. The homogeneous solution, $A^\mu_0$, obeys the vacuum wave equation and corresponds to incoming radiation. The second equation of motion, (\ref{LORENTZ}), is the covariant version of the Lorentz force law, the right-hand side being the Lorentz 4-force.

\section{Quantum Electrodynamics}

\subsection{Overview}

The following general rules apply when wants to quantise Maxwell theory:

\medskip

\noindent
\textbf{Rule 1:} Quantisation is required when the experimental resolution, hence the energy of a given probe, is sufficiently large such that photons become the relevant degrees of freedom. In this case, the classical wave picture has to be abandoned in favour of the particle (photon) picture. Historically, this has been noticed in black-body radiation (Planck's law), the photo-electric effect as explained by Einstein and the Compton effect, among many others.

\medskip
\noindent
\textbf{Rule 2:} Relativistic covariance has to be maintained throughout. 
\medskip

To implement these rules when quantising electrodynamics one might try to replace the particle equation of motion (\ref{LORENTZ}) by a relativistic version of the Schr\"odinger equation. This was attempted by Dirac with his celebrated Dirac equation. It turned out, however, that a formalism using single-particle wave functions obeying the Dirac equation does not work:  It does not allow for creation and annihilation processes where particle number ceases to be conserved. In particular, particle number does not commute with the generators of Lorentz boosts: different observers will thus disagree in their measurements of particle content! In more practical terms, once the localisation energy exceeds a threshold of $2mc^2$, the creation of an electron-positron pair of mass $2m$ will be possible. This has been seen in Klein's paradox \cite{Klein:1929zz}, where the relativistic scattering at a potential step seems to violate unitarity in that reflection and transmission probabilities do not add up to unity. The resolution of the paradox requires a `leakage' current due to the formation of pairs, a reaction channel that is absent in a single-particle formalism.    

Instead, one needs a relativistic many-body formalism based on `second quantisation', where both light and matter have to be described by quantum \emph{fields} with their modes quantised according to their statistics: Bose-Einstein for photons and Fermi-Dirac for electrons (and positrons). The result is quantum electrodynamics (QED), the first realistic example of a relativistic quantum field theory which is the only consistent unification of quantum mechanics and special relativity. The physical content of QED is encoded in its Lagrangian,
\be
  \mathcal{L}_\mathrm{QED} = \mathcal{L}_\mathrm{Maxwell} + \mathcal{L}_\mathrm{Dirac} 
  + \mathcal{L}_\mathrm{int} = -\frac{1}{4} F^2 + \bar{\psi} (i \hbar \slashed{\partial} - mc) \psi + e j \cdot A \; . 
  \label{QED}
\ee
Each of these terms may be symbolically associated with a Feynman diagram, 
\be 
   \includegraphics[scale=0.5]{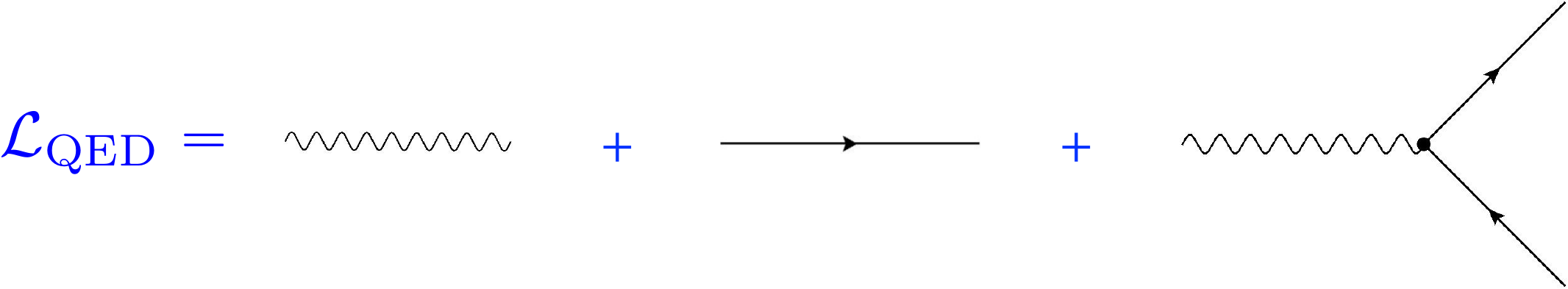}
\ee
the last term describing the fundamental QED vertex, the interaction between photons and Dirac particles (electrons, positrons). The latter are encoded in the electromagnetic or Dirac current, $j_\mu = \bar{\psi} \gamma_\mu \psi$ with $\gamma_\mu$ denoting the four gamma matrices required for a first-order relativistic wave equation. The form of the interaction is dictated by the principle of gauge invariance which implies `minimal substitution', i.e.\ the replacement of mechanical by canonical momenta or derivatives by gauge \emph{covariant} derivatives, $i \slashed{\partial} \to i \slashed{\partial} -e \slashed{A}$. (The Feynman-`slash' notation represents contraction with the gamma matrices, $\slashed{a} \equiv \gamma \cdot a$.)

Looking at the QED Lagrangian we can identify four parameters, $\hbar$, $c$, $m$ and $e$ corresponding to Planck's constant, the speed of light, the electron mass and charge, respectively. The first two of these, $\hbar$ and $c$, reflect the union of quantum mechanics and special relativity and can be set to unity without harm upon choosing natural units. This renders the elementary charge, $e$, dimensionless, so that the only dimensionful scale remaining is the mass $m$. Using dimensional analysis, one infers the important basic quantities listed in Table~\ref{tab:1}.

\renewcommand{\arraystretch}{1.3}

\begin{table}[h!]
\centering
\begin{tabular}{|l|l|l|}
\hline
quantity & formula & name \\
\hline \hline
energy scale & $E_0 \equiv mc^2$ & electron rest energy \\
\hline
length scale & $\lambdabar_e \equiv  \hbar/mc$ & electron Compton wave length \\
\hline
field strength & $e\mathcal{E}_S \equiv E_0/\lambdabar_e = m^2 c^3/\hbar$ & Sauter-Schwinger field \\
\hline
coupling & $\alpha = E_C/E_0 = e^2/4\pi \hbar c$ & fine structure constant \\
\hline
\end{tabular}
\caption{\label{tab:1}Important quantities derived from basic QED parameters}
\end{table}

The first entry, the electron rest energy, $E_0 = mc^2 \simeq 0.5$ MeV, provides the QED energy scale. Adopting natural units henceforth, $E_0 = m$, while its inverse defines the QED length scale, $\lambdabar_e = 1/m \simeq 400$ fm. Energy per length, thus $E_0/\lambdabar_e$, is a force, which defines the QED field strength, $\mathcal{E}_S = m^2/e = 1.3 \times 10^{18}$ V/m. This field magnitude is \emph{typical} for QED across distances of the order of a Compton wavelength. The challenge is to achieve such field strengths over macroscopic distances. Last but not least, the fine structure constant is defined as the ratio of the Coulomb energy between two electrons separated by a Compton wavelength and the electron rest energy, $\alpha = e^2/4\pi = 1/137 \ll 1$. The small value of this basic QED coupling guarantees that perturbation theory in $\alpha$ works well and yields highly accurate results. 

The reader may have noticed the subscript $S$ associated with the QED field strength. This is to flag the historical contributions of Sauter and Schwinger. The first introduced this quantity as early as 1931 in his tunnelling interpretation of the Klein paradox \cite{Sauter:1931zz}. Schwinger performed the first modern QED calculation of the pair creation rate, $R$, in the presence of a uniform electric field, with the famous result \cite{Schwinger:1951nm}:
\be
  R \sim E^2 \exp(-\pi \mathcal{E}_S/E) 
  = E^2 \exp(-\pi m^2/eE)\; . \label{SCHWINGER}
\ee
This rate is \emph{nonperturbative} in the coupling $e$ and represents a huge exponential suppression for field strengths $E \ll \mathcal{E}_S$.  Schwinger's rate (\ref{SCHWINGER}) can be made Lorentz invariant by employing the invariant fields (\ref{E.INV}) and (\ref{B.INV}) which leads to \cite{Nikishov:1969tt}
\be
  R \sim \mathcal{E}\mathcal{B} \, \coth(\pi \mathcal{B}/\mathcal{E}) 
  \, \exp(-\pi m^2 /e\mathcal{E}) \; .
\ee

\subsection{Application: Compton Scattering}

\begin{figure}[h!]
\begin{center}
\includegraphics[scale=0.6]{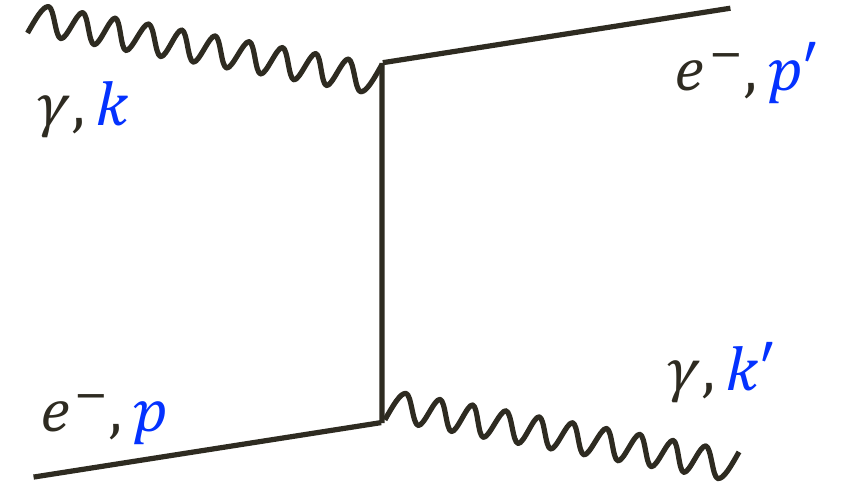}
\end{center}
\caption{\label{fig:COMPTON} Feynman diagram for Compton scattering with 4-momentum assignments.}
\end{figure}

As an important application we will consider the elementary process of Compton scattering where an electron and a photon scatter off each other. This was first discovered by Compton in 1923 \cite{Compton:1923zz} who shot X-rays at electrons at rest and observed a red-shift of the X-ray photons corresponding to an energy-momentum transfer from the photons to the electrons. The process is represented by the Feynman diagram of Fig.~\ref{fig:COMPTON}.

It shows an electron of 4-momentum $p$ colliding with a photon of 4-momentum $k$ exchanging energy and momentum such that the particles end up with 4-momenta $p'$ and $k'$, respectively. This description is relativistically covariant. Any choice of frame is encoded in the parametrisation of the initial 4-momenta. Whatever this choice, one always has energy momentum conservation, 
\be 
  p + k = p' + k' \; .
\ee
One can actually do `better' and introduce the Mandelstam invariants \cite{Mandelstam:1958xc} which have the same value in any frame,
\bea
  s &=& (k+p)^2 \equiv m^2(1 + 2 \eta) \; , \label{MANDELSTAM.S}\\
  t &=& (k' -k)^2 = - 2 k \cdot k' \; , \\
  u &=& (p - k')^2 = m^2 (1 - 2 \eta') \; .
\eea
Here we have also introduced the dimensionless energy variables
\be
  \eta := \frac{\hbar k \cdot p}{(mc)^2} \; , \quad \eta' := 
  \frac{\hbar k \cdot p'}{(mc)^2} \; , \label{ETA}
\ee
which measure the electron momentum projection along the photon direction $k$ in units of $(mc)^2$. (We have temporarily reinstated $\hbar$ and $c$ to flag that these are relativistic quantum variables.)

\begin{figure}[h!]
\begin{center}
\includegraphics[scale=0.6]{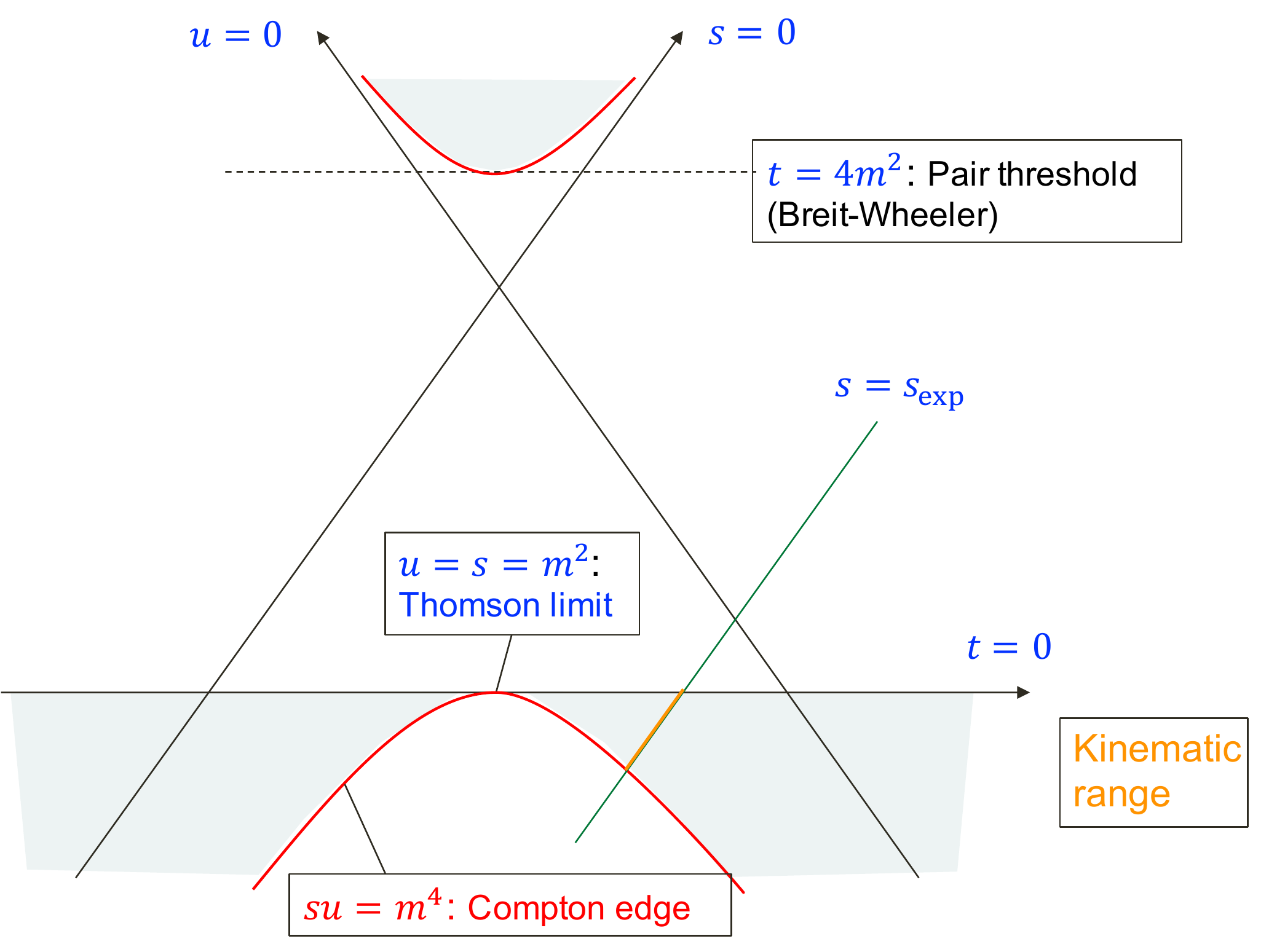} 
\caption{\label{fig:MANDELSTAM1} Mandelstam plot for Compton scattering. (The pair threshold is not drawn to scale to save space.)}
\end{center}
\end{figure}

The most important quantity is arguably $s$, defined in (\ref{MANDELSTAM.S}), which represents the total energy (squared) of the particles in the centre-of-mass frame. In other words, this is the available energy budget for the process expressed in a frame independent manner. The Mandelstam variables are not independent but obey the constraint 
\be \label{MANDELSTAM.CONSTRAINT}
  s + t + u = 2 m^2 \; , 
\ee
which follows directly from the mass shell conditions,
\be 
  p^2 = p'^2 = m^2 \; , \quad k^2 = k'^2 = 0 \; .
\ee
Using (\ref{MANDELSTAM.CONSTRAINT}) the invariant kinematics of the Compton process can be illustrated with a Mandelstam plot as depicted in Fig.~\ref{fig:MANDELSTAM1}. 

For any given point in the plane, the Mandelstam variables are given by the orthogonal distances to the axes $s, t, u = 0$, which form an equilateral triangle of height $2m^2$. The allowed kinematic range for Compton scattering is given by the shaded area enclosed by the (horizontal) axis $t=0$  and the parabola $su =m^4$ (in red) defining the \emph{Compton edge}. It corresponds to the maximal momentum transfer  
\be
  -t = (s - m^2)^2/s = 4 (k \cdot p)^2/s = 
  \frac{4\eta^2 m^2 }{ 1 + 2\eta} \; ,
\ee
or back-scattering kinematics, where the 3-momenta of incoming and outgoing photons have opposite directions. The classical (Thomson) limit is located at the vertex of the parabola where $s = u = m^2$ and $t=0$ (no momentum transfer or recoil). The kinematic regions to the left and right of this point are referred to as the $u$ and $s$ channel, respectively. The $t$ channel, on the other hand, is given by the upper parabolic region where $ t \ge 4 m^2$, the pair creation threshold (not drawn to scale to save space). This is the kinematic region for the crossed (pair creation) process, $\gamma + \gamma \to e^+ + e^-$, first predicted by Breit and Wheeler in 1934 \cite{Breit:1934zz}. A version of this process with the incoming photons still slightly virtual or off-shell ($k^2 \ne 0$) has recently been observed for the first time \cite{STAR:2019wlg}.

\begin{figure}[h]
\begin{center}
\includegraphics[scale=0.7]{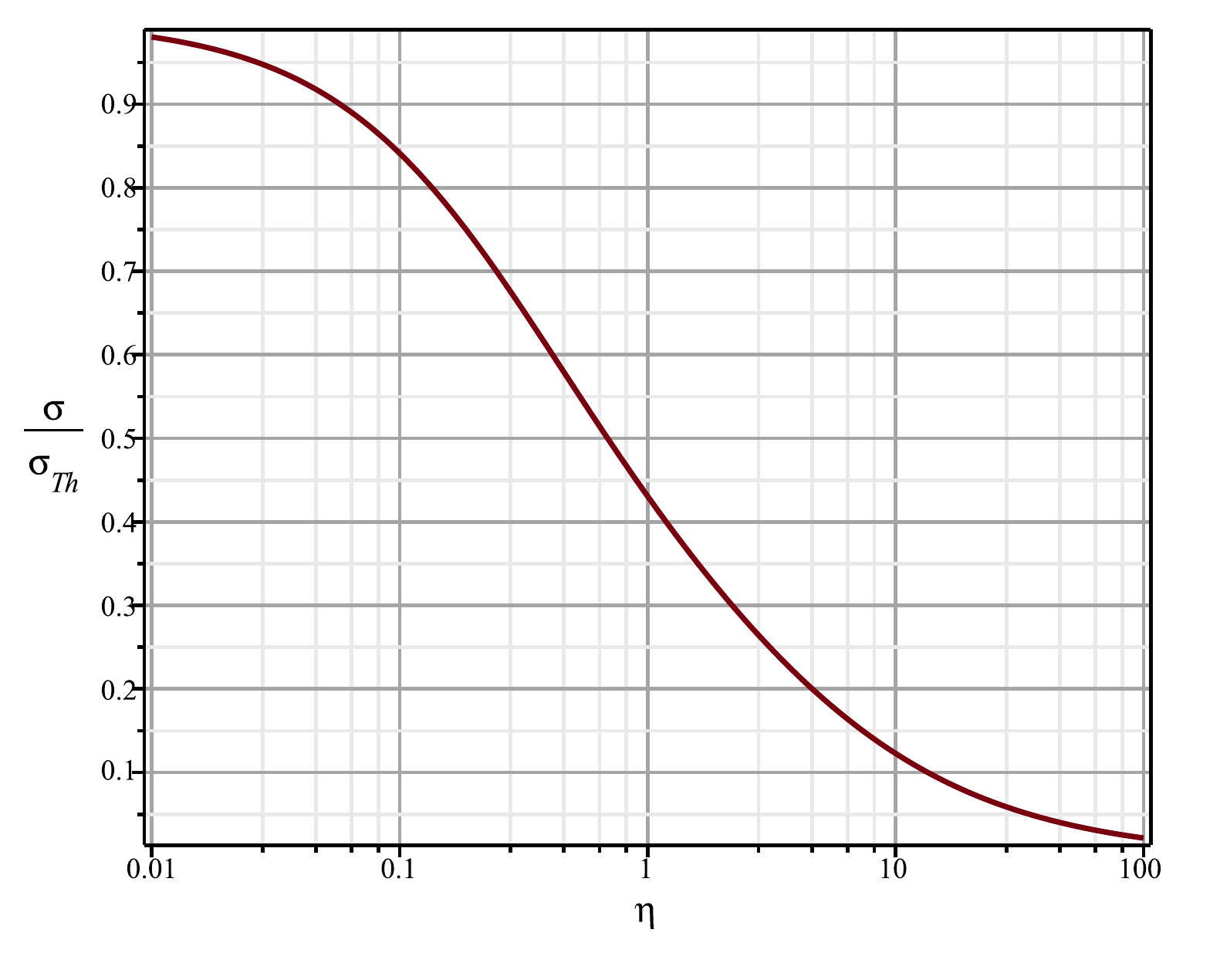} 
\caption{\label{fig:KLEIN-NISHINA}Klein-Nishina cross section for QED Compton scattering in units of the classical Thomson cross section.}
\end{center}
\end{figure}

The total cross section for Compton scattering, first calculated by Klein and Nishina in 1928, is depicted in Fig.~\ref{fig:KLEIN-NISHINA} as a function of $\eta$ in units of the classical Thomson cross section, 
\be
  \sigma_\mathrm{Th} = \frac{8 \pi}{3} \frac{\alpha^2}{m^2} \; ,
\ee
which is energy independent. Quantum effects thus lead to a significant reduction of the cross section at high energy ($\eta \gg 1$). 

\section{High-Intensity Quantum Electrodynamics}

\subsection{Introduction}

It is customary in physics to investigate the response of a system to an applied external field. Well known examples include spin systems or superconductors in the presence of an external magnetic field. In the context of QED, the study of external field problems goes back at least to the work of Heisenberg and Euler on vacuum polarisation \cite{Heisenberg:1936nmg}. They assumed the external electric or magnetic fields to be constant in space and time. 

Here we want to analyse how QED gets modified under the influence of a high-intensity laser field. The simplest model for such a field is a plane wave characterised by a wave vector $k$ that is light-like or null, $k^2 = 0$. The associated 4-potential will only depend on the invariant phase variable, $k \cdot x$, and hence obeys the free wave equation, $\Box A_\mu = 0$, while the field tensor is null as well. This means that the field invariants (\ref{S}) and (\ref{P}) vanish,
\be
  \mathcal{S} = \mathcal{P} = 0 \; , 
\ee 
such that electric and magnetic fields, $\vc{E}$ and $\vc{B}$, are orthogonal and of equal magnitude\footnote{Of course, this plane wave model is an over-simplification as it does not describe a focussed beam.}. Null fields are hence rather elusive as has been known already to Heisenberg who noted in 1934 that a single plane wave cannot polarise the vacuum. One needs at least two of them brought into collision to ensure a non-vanishing energy density \cite{Heisenberg:1934pza}. This has been confirmed by a modern QED calculation due to Schwinger who showed that plane waves lead to a vanishing effective action, hence the absence of vacuum polarisation \cite{Schwinger:1948yj}.

Thus, in order to define non-vanishing invariants, the (plane wave) fields alone are not sufficient, and one has to employ a probe particle, say an electron, of momentum $p = mu$. One can then define the laser frequency and the energy density `seen' by this probe (i.e.\ as measured in the instantaneous rest frame of the electron), namely
\bea
  \omega_L &:=& k \cdot u \; , \\
  \mathcal{E}_0^2 &:=& (u, F^2, u)/c^2 \; .
\eea
These can be made dimensionless upon introducing the parameters
\bea
  \eta &:=& \frac{\hbar \omega_L}{mc^2} \; , \\
  \chi &:=& \frac{e\mathcal{E}_0 \lambdabar_e}{mc^2} \; ,
\eea
referred to as the quantum energy and the quantum nonlinearity parameters, respectively. (Note that $\eta$ formally coincides with the definition (\ref{ETA}), but with $k$ now denoting the laser momentum.) Dividing the two one obtains the classical nonlinearity parameter,
\be 
  a_0 := \chi/\eta = \frac{e\mathcal{E}_0 \lambdabar_L}{mc^2} \; ,
\ee
defined in terms of  the field magnitude $\mathcal{E}_0$ `seen' by the probe and the reduced laser wavelength,  $\lambdabar_L = c/\omega_L$. Thus, $a_0$ is the dimensionless laser field amplitude (sometimes denoted as $\xi$). The parameters above have to be added to those of standard QED which obviously enlarges the parameter space to be studied. One is particularly interested in the intensity, hence $a_0$, dependence at a given value of $\eta$, i.e.\ the centre-of-mass energy of the combined laser-electron system, cf.\ (\ref{MANDELSTAM.S}) and the application further below.

\subsection{Formalism}

\begin{figure}[h!]
\begin{center}
\includegraphics[scale=0.5]{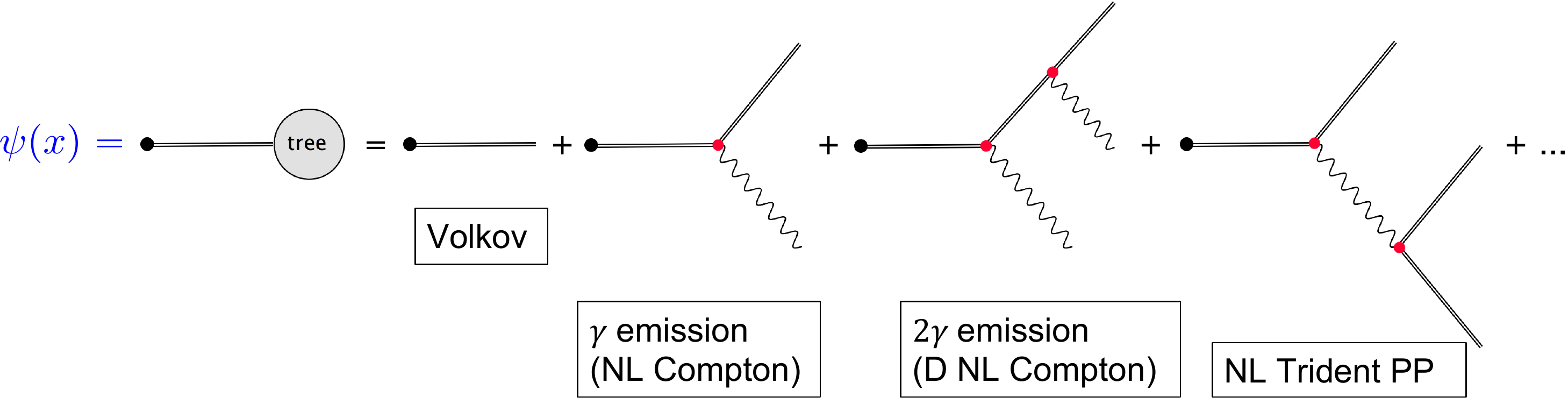}
\end{center}
\caption{\label{fig:SOLUTION.PSI} Solution of the electron equation of motion (\ref{EOM.PSI.BG}). The leading order on the right represents the Volkov solutions while higher orders correspond to particle emission (a.k.a.\ nonlinear or double nonlinear Compton scattering) or pair production (PP) processes.}
\end{figure}

\begin{figure}[h]
\begin{center}
\includegraphics[scale=0.5]{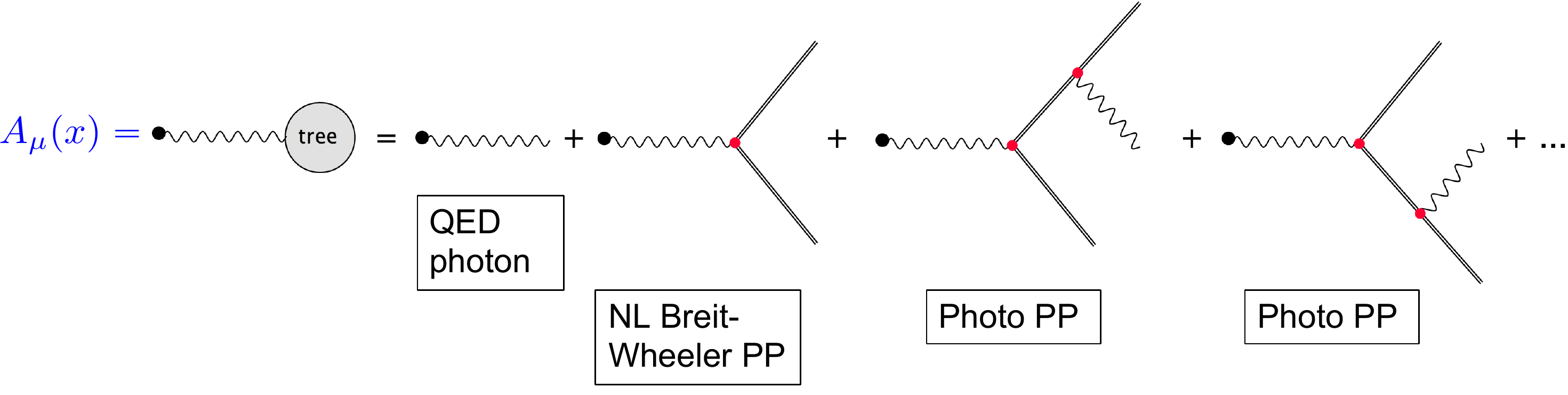}
\end{center}
\caption{\label{fig:SOLUTION.A} Solution of the photon equation of motion  (\ref{EOM.A}). The leading order on the right represents a free QED photon while higher orders correspond to particle emission and/or pair production processes.}
\end{figure}

Before discussing an application let us have a brief look at the basic formalism of high-intensity QED.  In the QED Lagrangian, one splits the 4-potential into a plane wave \emph{background} field, $\bar{A}= \bar{A} (k\cdot x)$,  and a fluctuating photon, $A$. The background is null and obeys the vacuum wave equation, $\Box \bar{A} = 0$. Its kinetic term, the invariant $\bar{\mathcal{S}}$, vanishes while the mixed term, $F \bar{F}$, is a total derivative that does not contribute to the action. Hence, the background field $\bar{A}$ enters the Lagrangian solely through its coupling to the matter current, $\bar{\mathcal{L}}_\mathrm{int} = e\bar{A} \cdot j$. Setting $e\bar{A} =: a_0 a $, the Dirac equation thus becomes
\be
  (i \slashed{\partial} - m - a_0 \slashed{a}) \psi = e \slashed{A} \psi \; , \label{EOM.PSI.BG} \\
\ee
while the equation for the QED photon, $A$, remains unchanged,
\be 
  \Box A_\mu = e \bar{\psi} \gamma_\mu \psi \; . \label{EOM.A}
\ee
The important thing to note is that we are now dealing with \emph{two} couplings, $e$ and $a_0$. While $e < 1$ (and in particular $\alpha = e^2/4\pi \ll 1$), the coupling $a_0$ can be much larger than unity. In other words, while perturbation theory in $e$ (or, effectively, $\alpha$) makes sense, we need to treat $a_0$ non-perturbatively, trying to sum up all orders in $a_0$.  Thus, the right-hand sides of (\ref{EOM.PSI.BG}) and (\ref{EOM.A}) can be considered small but the zeroth-order approximation to (\ref{EOM.PSI.BG}), obtained by setting $e=0$ on the right, involves all orders in $a_0$ on the left. Fortunately, there is an exact zeroth-order solution due to (and named after) Volkov \cite{Wolkow:1935zz}. Perturbing on top of a background that is treated exactly  corresponds to working in the quantum mechanical Furry picture \cite{Furry:1951zz}. Rather than writing down a formula, we will proceed graphically by employing the fact that the solutions of the classical equations of motion (\ref{EOM.PSI.BG}) and (\ref{EOM.A}) are found by summing all tree diagrams formed from the leading-order solutions. Representing the Volkov solution by a double line, the solutions to order  $e^2$ are depicted in Fig.s \ref{fig:SOLUTION.PSI} and \ref{fig:SOLUTION.A}. 

The first high-intensity QED calculation was done by Reiss who considered nonlinear Breit-Wheeler pair production in 1962 \cite{Reiss:1962}. This was soon followed by extensive work on photon emission and pair production due to Nikishov, Ritus, Narozhny and others in the (then) Soviet Union \cite{Nikishov:1964zza,Nikishov:1964zz,Narozhny:1965,Goldman:1964aka} large parts of which are reviewed in \cite{Ritus:1985}.  At the same time, important and lucid contributions analysing (nonlinear) Compton scattering at high intensity were made by Brown and, in particular, Kibble \cite{Brown:1964zzb,Kibble:1965zza,Kibble:1966zz,Kibble:1966zza}.

The discussion of higher-order processes has included processes such as double nonlinear Compton scattering \cite{Lotstedt:2009zz,Lotstedt:2009zza,Seipt:2012tn,Mackenroth:2012rb,King:2014wfa,Dinu:2018efz} and the nonlinear trident process \cite{Baier:1972,Ritus:1972,Morozov:1977vv,Bula:1997eh,Hu:2010ye,Ilderton:2010wr,King:2013osa,Dinu:2017uoj,Mackenroth:2018smh}. These studies generalise their linear pendants which already are nontrivial due to the three-particle final states. (See e.g.\ Ch.~11 of the classic QED text by Jauch and Rohrlich \cite{Jauch:1976ava} for an overview and early references.) For the nonlinear processes, the current state of the art is presented in \cite{Dinu:2019wdw,Torgrimsson:2020mto} (and references therein). Photo-pair production (the two-vertex processes in Fig.~5) was first considered by means of the optical theorem, i.e.\ by cutting the relevant loop diagram \cite{Morozov:1977vv}. Recently, it has been revisited under the name of `photon trident' using modern techniques \cite{Torgrimsson:2020mto}. 

\subsection{Application: Nonlinear Compton Scattering}

A natural question to ask is about the modifications to Compton scattering if one of the incoming photons is replaced by a laser beam, i.e.\ a coherent superposition of optical-frequency (hence low-energy) photons. An alternative point of view is to regard this process as the emission of a photon by an electron dressed by, and exchanging 4-momentum with, the laser field (the one-vertex process in Fig.~\ref{fig:SOLUTION.PSI}).  

\begin{figure}[h!]
\begin{center}
\includegraphics[scale=0.6]{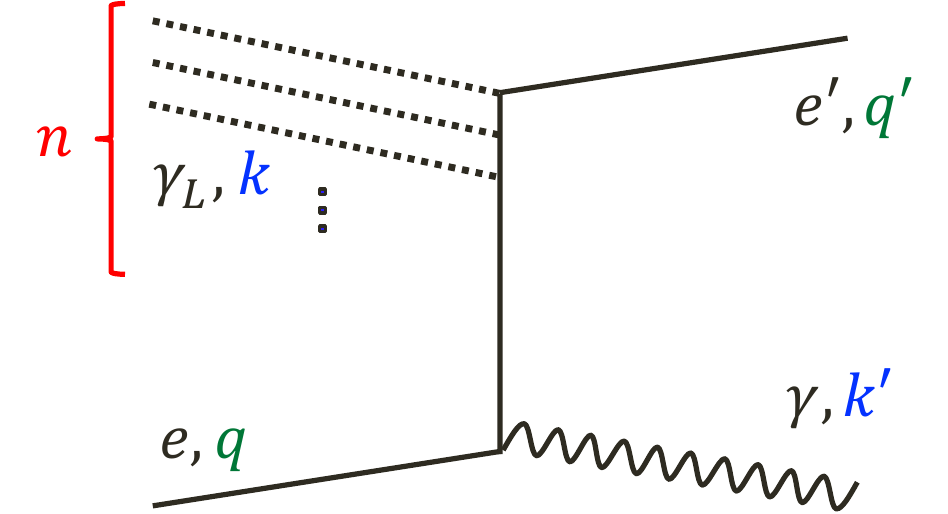}
\end{center}
\caption{\label{fig:NLC} Feynman diagram for nonlinear Compton scattering with 4-momentum assignments.}
\end{figure}

The process in question replaces the Feynman diagram of Fig.~\ref{fig:COMPTON} by a \emph{sum} over the diagrams depicted in Fig~\ref{fig:NLC}.

\begin{figure}[h]
\begin{center}
\includegraphics[scale=0.6]{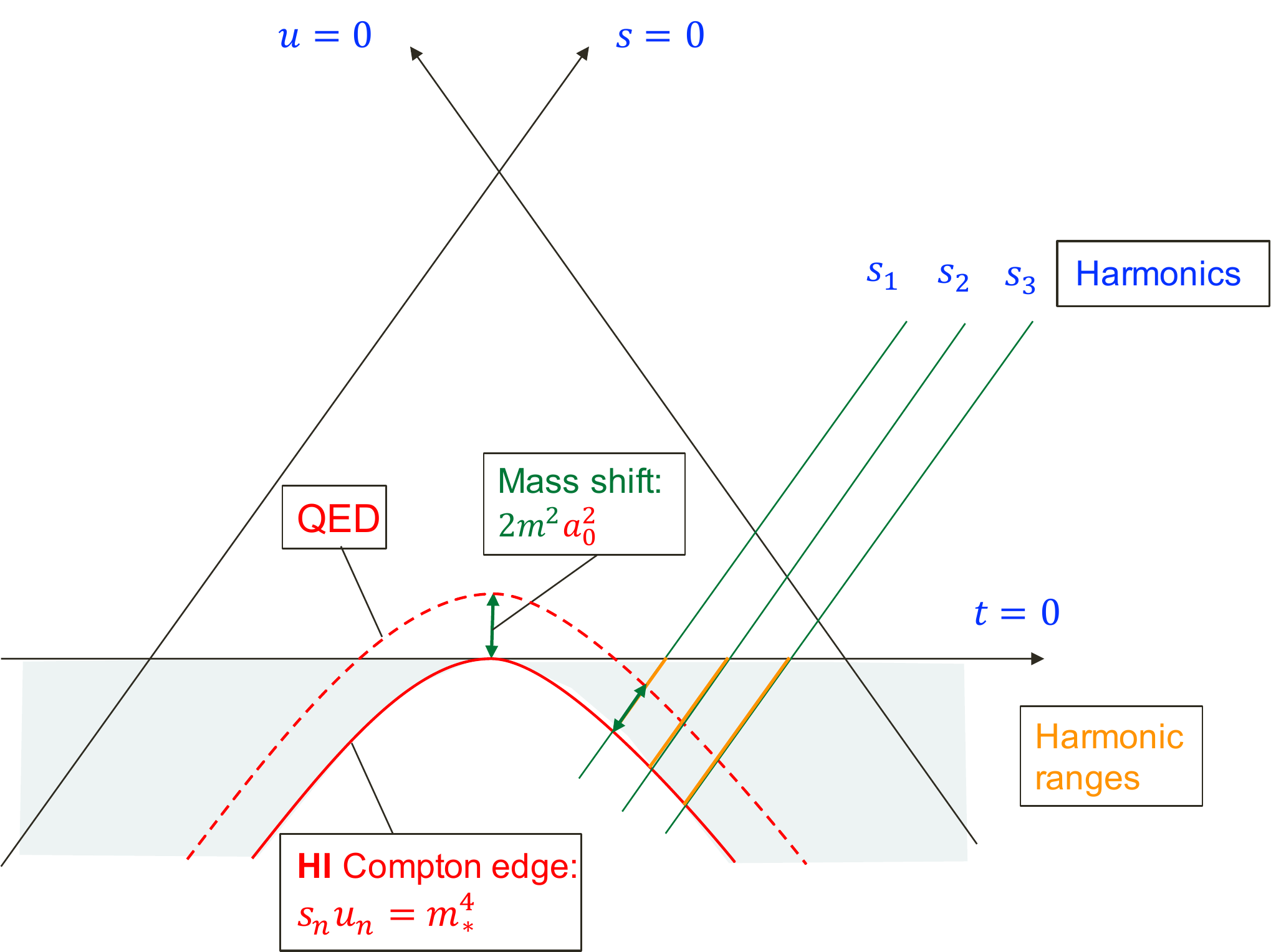}
\end{center}
\caption{\label{fig:MANDELSTAM2} Mandelstam plot for nonlinear Compton scattering. Compared to standard QED, one should note the intensity dependent mass shift and the appearance of harmonics, each with their own kinematic ranges.}
\end{figure}

The Feynman graph corresponds to the $n$-photon process $e + n \gamma_L \to e' + \gamma$. As any number $n$ of laser photons may be involved, the total amplitude is obtained by summing over all $n$. Formulae can be looked up in the original publications  \cite{Nikishov:1964zza,Nikishov:1964zz,Narozhny:1965} or in the textbook by Landau and Lifshitz \cite{Berestetskii:1982qgu}. We will not need the details with one exception, the appearance of the \emph{quasi-momenta}, 
\be
  q  := p +  \frac{a_0^2}{2\eta} k \; , \quad  q' := p' + \frac{a_0^2}{2\eta'} k \; ,
\ee
in the energy-momentum balance
\be 
  q + nk = q' = k' \; .
\ee
The quasi-momenta are thus intensity dependent, which is reflected in the altered mass-shell condition,
\be 
  q^2 = q'^2 = m^2 (1 + a_0^2) =: m_*^2 \; ,
\ee
The mass-shift, $m \to m_*$, was first discovered by Sengupta \cite{Sengupta:1952} and has been extensively discussed by Kibble \cite{Kibble:1965zza,Kibble:1966zz,Kibble:1966zza} so that $m_*$ also goes under the name of Kibble mass. The effects caused by this modified mass can again be nicely illustrated in terms of a Mandelstam plot \cite{Harvey:2009ry}. To this end, one introduces the modified Mandelstam variables,
\bea
  s_n &=& (nk + q)^2 = m^2 (1 + a_0^2 + 2n \eta) \; , \\
  t_n &=& (k' - nk)^2 = -2n k \cdot k' \; , \\
  u_n &=& (nk - q')^2 = m^2 (1 + a_0^2 - 2n \eta') \; .
\eea 
These obey the sum rule 
\be 
  s_n + t_n + u_n = 2 m_*^2 \; , 
\ee
which is independent of the photon number $n$, hence the same for all processes represented by Fig.~\ref{fig:NLC}. Crucially, this alters the height of the unilateral triangle in the Mandelstam plot which thus serves as an invariant signature for the mass shift, see Fig.~\ref{fig:MANDELSTAM2}. The $n$ dependence of the Mandelstam variables corresponds to the appearance of higher harmonics, labelled by $s_n$, each with its own kinematic range (highlighted in orange in Fig.~\ref{fig:MANDELSTAM2}).

\begin{figure}[h]
\begin{center}
\includegraphics[scale=0.6]{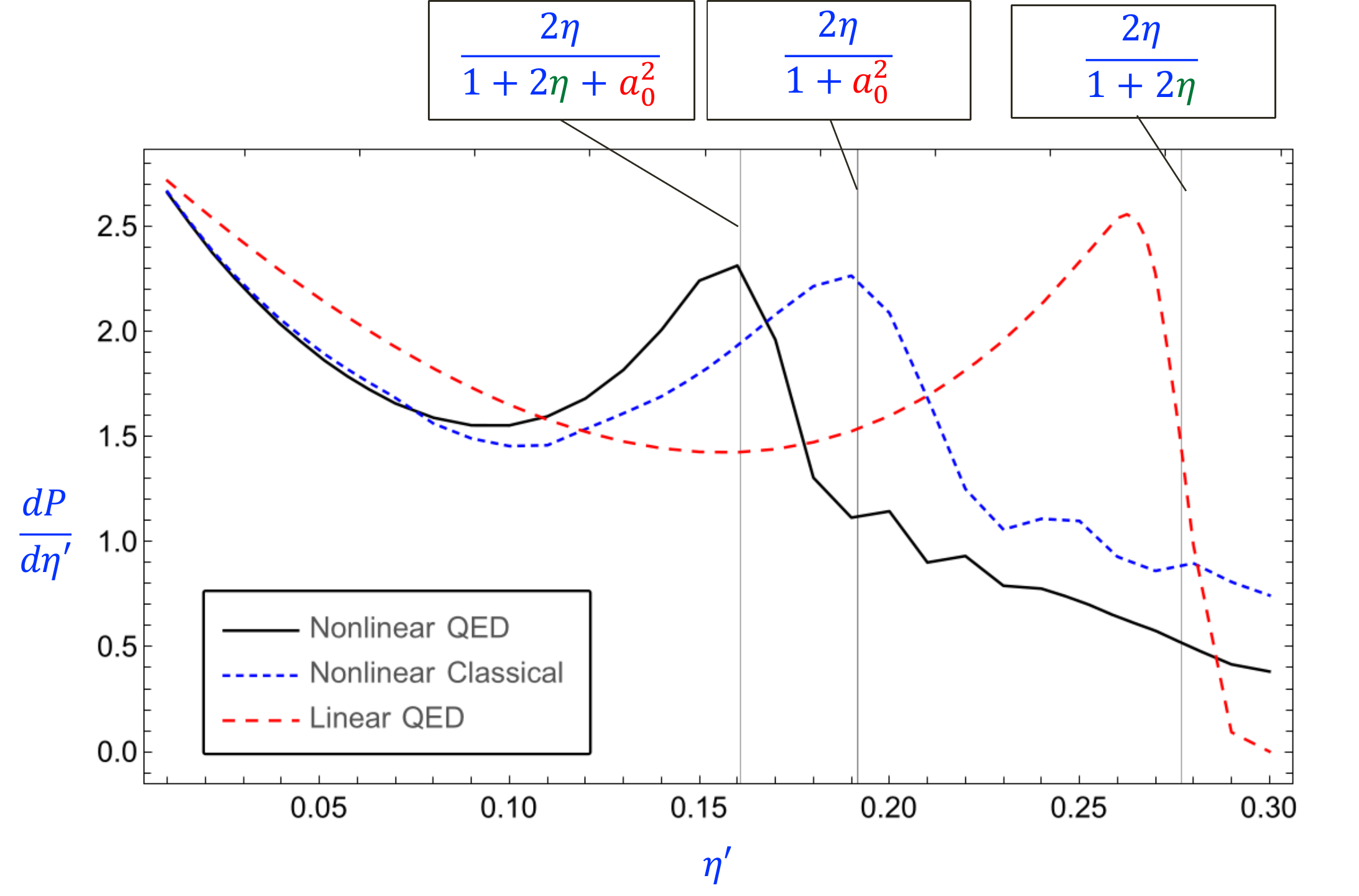}
\end{center}
\caption{\label{fig:SPECTRUM} Scattered photon spectra for electron laser scattering (adapted from \cite{Abramowicz:2021zja}). Vertical lines demarcate the kinematic edges for QED Compton scattering (red-dashed), nonlinear Thomson scattering (blue-dashed) and nonlinear Compton scattering (black).}
\end{figure}

Compared to linear Compton scattering (standard QED), the range in Mandelstam-$t$ gets enlarged corresponding to a larger momentum transfer. This implies a red-shift (blue-shift) of the fundamental, $n=1$,  Compton edge in the scattered photon (electron) spectra. Fig.~\ref{fig:SPECTRUM} compares the photon spectra for linear Compton as well as nonlinear Thomson and Compton scattering employing the parameters of the LUXE experiment \cite{Abramowicz:2021zja}. On top, we have listed the formulae for the different Compton edges. One can clearly distinguish between classical intensity effects (parametrised by $a_0$) and quantum effects (parametrised by $\eta$).

\begin{figure}[h]
\begin{center}
\includegraphics[scale=0.6]{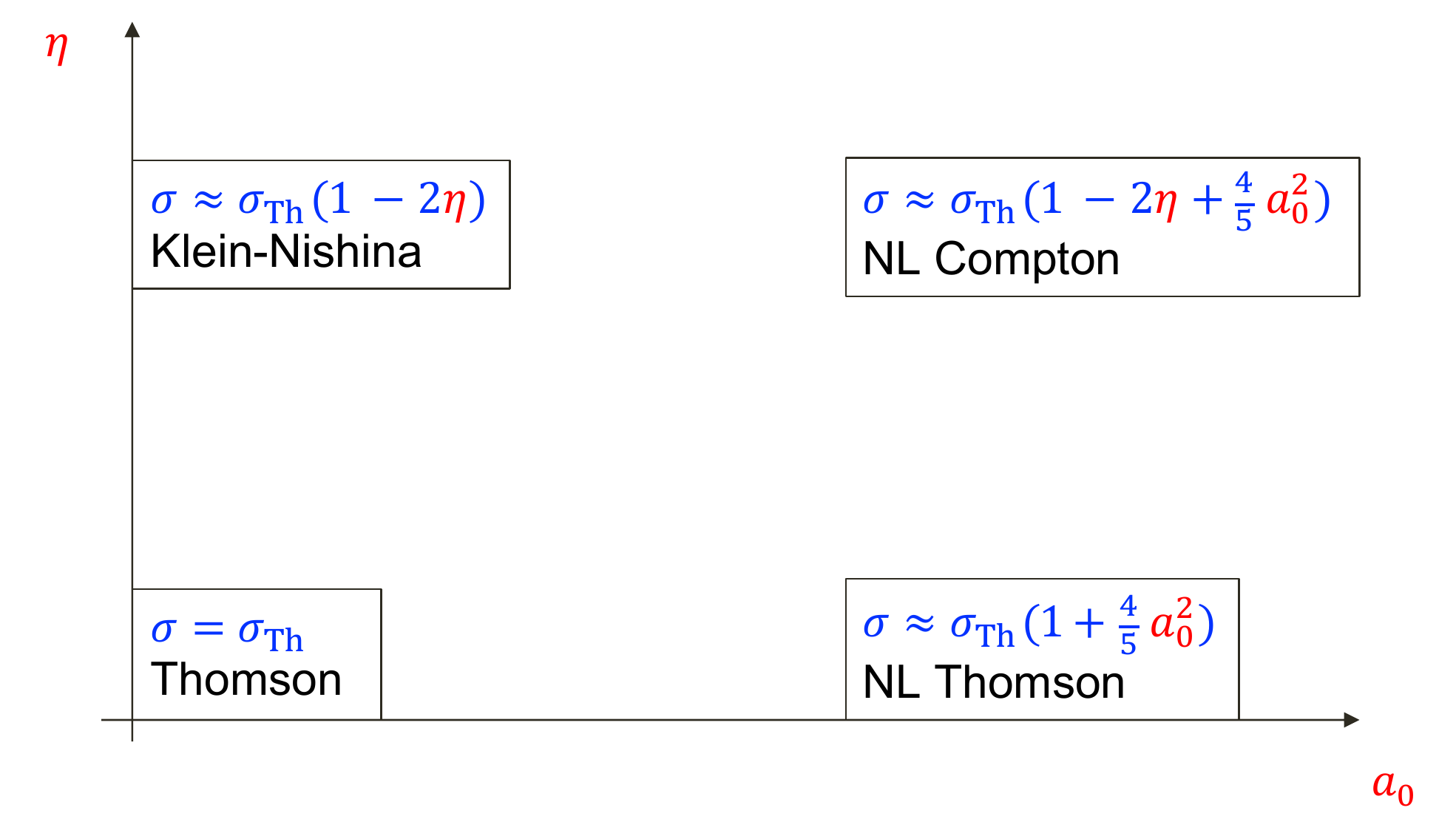}
\end{center}
\caption{\label{fig:TOTAL-XSECTION} Leading-order corrections to the Thomson cross section induced by intensity and quantum effects.}
\end{figure}

As before, we can compare the total cross sections. The formulae are quite messy, so we just have a look at the corrections to leading-order in $a_0$ and $\eta$ \cite{Heinzl:2013gja}. Fig.~\ref{fig:TOTAL-XSECTION} provides an overview of the results for the different regions of parameter space (in the $a_0 \eta$ plane). 

The corrections are in one-to-one correspondence with the three spectra of  Fig.~\ref{fig:SPECTRUM}: The Klein-Nishina cross-section for (linear) QED Compton scattering was already shown in Fig.~\ref{fig:KLEIN-NISHINA}. For nonlinear (high-intensity) QED it gets replaced by the nonlinear Compton cross section which has the nonlinear Thomson cross section as its classical limit ($\eta \to 0$).

\subsection{Teaser: Radiation Reaction}

It turns out that there is a classical correction to the Thomson cross section that is missing in Fig.~\ref{fig:TOTAL-XSECTION}, namely that induced by radiation reaction. It was first calculated by Dirac in 1938 when he posed his relativistic equation of motion \cite{Dirac:1938nz} that replaces (\ref{LORENTZ}). Taking into account the effects of radiation through elimination of the Li\'enard-Wiechert potentials he obtained what is now called the Lorentz-Abraham-Dirac equation,
\be
  m \ddot{x} = e F \dot{x} + F_\mathrm{rad} \; . \label{LAD}
\ee
We have refrained from explicitly writing down the correction term, $F_\mathrm{rad}$. It suffices to say that it is proportional to the third (proper) time derivative of position $x$ (which causes problems in its own right) and a time parameter,
\be
  \tau_0 := \sfrac{2}{3} \alpha \lambdabar_e \simeq 2 \; \mbox{fm}/c \; .
\ee
Note that the powers of $\hbar$ cancel in this expression. From his equation (\ref{LAD}), Dirac finds that the Thomson cross section should be replaced by
\be
  \sigma_\mathrm{RR} \equiv \frac{\sigma_\mathrm{Th}}{1 + 2 \, k \cdot u \, \tau_0} = 
  \frac{\sigma_\mathrm{Th}}{1 + \sfrac{2}{3} \alpha \eta}
  \simeq \sigma_\mathrm{Th} \left(1 - \sfrac{2}{3} \alpha \eta \right) \; .
\ee
Again, factors of $\hbar$ have all cancelled. It is an unsolved question how this cross section arises within QED. Recent investigations suggest \cite{Heinzl:2021mji,Ekman:2021eqc,Torgrimsson:2021zob} that it may result from an all-orders resummation of an appropriate class of Feynman diagrams including loops.  This resummation should yield the summed geometric series (in $ \sfrac{2}{3} \alpha \eta$) represented by Dirac's cross section, $\sigma_\mathrm{RR}$.   

\section{Conclusion}

In the course of the last decade or so, high-intensity QED has become a mature field of research with new dedicated experiments close to realisation so that the findings reported in \cite{E144:1996enr,Burke:1997ew,Bamber:1999zt,Cole:2017zca,Poder:2017dpw} will soon be confirmed and/or improved upon. As a result, it is not possible to do justice to the whole field in a half-hour tutorial session. The required focussing on just a few aspects has hence led to a number of omissions which include: laser induced pair production and related processes, phenomena related to vacuum polarisation and light-by-light scattering as well as higher-order processes and fundamental questions related to them such as the breakdown of strong-field perturbation theory. One also needs to address the limitations of the approximations made, in particular the plane wave and external field approximations. The plane wave model cannot describe focussed beams, while an external field by definition is blind to back-reaction phenomena caused by that very field. Space and time limitations have also led to a focus on theory, for which the author apologises to his experimental colleagues, in particular to those involved in the planned LUXE experiment.     

There are a number of recent reviews which cover the omissions made here and include references to the original works \cite{DiPiazza:2011tq,Narozhny:2015vsb,King:2015tba,Seipt:2017ckc,Blackburn:2019rfv,Karbstein:2019oej,Zhang:2020lxl,Gonoskov:2021hwf,Fedotov:2022ely}. The reader is encouraged to find further information and inspiration there. 

\subsection*{Acknowledgements} 

The author thanks Alex Robinson for the invitation to give this presentation. He is indebted to Ben King for a careful reading of the manuscript and a number of useful comments.


\end{document}